# The Use of Unlabeled Data in Predictive Modeling

**Feng Liang, Sayan Mukherjee and Mike West**

*Abstract.* The incorporation of *unlabeled data* in regression and classification analysis is an increasing focus of the applied statistics and machine learning literatures, with a number of recent examples demonstrating the potential for unlabeled data to contribute to improved predictive accuracy. The statistical basis for this *semisupervised* analysis does not appear to have been well delineated; as a result, the underlying theory and rationale may be underappreciated, especially by non-statisticians. There is also room for statisticians to become more fully engaged in the vigorous research in this important area of intersection of the statistical and computer sciences. Much of the theoretical work in the literature has focused, for example, on geometric and structural properties of the unlabeled data in the context of particular algorithms, rather than probabilistic and statistical questions. This paper overviews the fundamental statistical foundations for predictive modeling and the general questions associated with unlabeled data, highlighting the relevance of venerable concepts of sampling design and prior specification. This theory, illustrated with a series of central illustrative examples and two substantial real data analyses, shows precisely when, why and how unlabeled data matter.

## 1. INTRODUCTION

Recent interest in the use of so-called unlabeled data in problems of prediction in the machine learning community has generated a growing awareness of the potential for incorporation of ancillary design data in classification and regression problems (Bennett and Demiriz, 1999; Blum and Mitchell, 1998; Joachims, 1999; Szummer and Jaakkola, 2002; Zhu, Ghahramani and Lafferty, 2003; Belkin, Niyogi and Sindhwani, 2004). This use of unlabeled data is often referred to as semisupervised learning. Mainstream probabilistic thinking is relatively underrepresented in this active and exciting literature, and the theoretical underpinnings of algorithms that exploit unlabeled data have received scant attention from statistical scientists. Much of the activity is algorithmic and applied. Machine learning examples are typically presented case-by-case, with the semisupervised analysis usually based on modifications of (fully supervised) optimization algorithms for classification or regression prediction, and with the introduction of additional components of objective functions that tie in unlabeled samples. Arguments for these additional components are made using a combination of structural and intuitive arguments, including, most recently, asymptotic arguments on the convergence of operators on manifolds (Belkin and Niyogi, 2005; Coifman et al., 2005a, b). There has been some work addressing the theoretical aspects of unlabeled data (Castelli and Cover, 1995; Seeger, 2000; Cozman and Cohen, 2002; Ando and Zhang, 2005) in specific contexts. However, in general, the foundation and rationale for understanding the relevance, and likely effectiveness, of unlabeled data are still not well understood.

*Feng Liang is Assistant Professor, Department of Statistical Science, Duke University, Durham, North Carolina 27708, USA. She is on leave at the Department of Statistics, University of Illinois at Urbana-Champaign, 725 South Wright Street, Champaign, Illinois 61820, USA e-mail: feng@stat.duke.edu; liangf@uiuc.edu. Sayan Mukherjee is Assistant Professor, Department of Statistical Science and Institute for Genome Sciences and Policy, Duke University, Durham, North Carolina 27708, USA. Mike West is Arts and Sciences Professor, Department of Statistical Science, Duke University, Durham, North Carolina 27708, USA.*

For currently active application areas and also to underlie growth and development of the unlabeled data methodology long-term, it is critical that the underlying theoretical basis for the use of unlabeled data is delineated and more broadly understood among statistical and computational scientists. Our goal here is to promote broader awareness and interest among statisticians of the nature and importance of this area. We do this by outlining the conceptual and theoretical bases for the "when, why and how" in regard to the use of unlabeled data, and through a complementary series of illustrations in central statistical modeling contexts as well as empirical examples in two substantive data analyses.

Beginning in Section 2 with an articulation of the basic model framework and discussion of fundamental issues of sampling and design, we discuss the underlying conceptual and theoretical basis for using unlabeled data. This is developed in the Bayesian framework for prediction, in which implications for the incorporation, or otherwise, of unlabeled data in prediction problems becomes transparent. Section 3 provides concrete, illuminating examples in a series of common statistical models. This includes examples in regression, prediction using multivariate normal mixture models, and standard mixture-based classification and discrimination. These are key contexts that connect intimately with some of the major areas of interest in machine learning, and contexts in which the relevance of unlabeled data is perhaps most transparent and intuitive. These examples serve to highlight the relevance of unlabeled data in standard, central areas of statistics. Section 4 overviews and exemplifies the issues in a class of latent factor regression models, with an empirical illustration in analysis of a benchmark data set of handwritten digit classification. Section 5 concerns our final important context, that of kernel regression; here we link statistical and machine learning approaches, illustrate the theoretical basis for the use of unlabeled data and provide a further empirical study in discrimination of cancer and normal tissue samples based on gene expression data. We close with summary comments in Section 6.

## 2. GENERAL FRAMEWORK

### 2.1 Context, Goals and Models

Interest lies in aspects of the joint distribution of two random quantities, $(y, x)$, and the core prediction problem concerns statements about future values of $y$ based on observing the corresponding $x$. Both $x$ and $y$ may be multivariate, in general. In standard regression problems, $y$ is a continuous or discrete univariate response; in problems of classification, $y$ is discrete, often binary. Using $p(\cdot)$ as generic notation for probability density functions, all inference problems require understanding aspects of the joint density $p(y, x | \star)$, where $\star$ denotes all parameters—to be described below in context—that are needed to fully specify the joint density.

The fundamental problem of prediction—whether it be couched in terms of regression estimation or classification—is framed as follows: at a specified "future" value of $x$, make statements about the corresponding value of $y$. Using $*$ to denote future values of interest, this implies a directional focus: we want to understand and evaluate, or estimate, $p(y_* | x_*, D)$ based on all available data and information $D$.

Statistical models structure the problem in terms of parameters (which may be infinite-dimensional in nonparametric models) that represent all uncertain aspects of the joint probability distribution for $(y, x)$. By way of notation, the dominant and generally (our) preferred specification of the joint density is

$$(1) \qquad p(y, x | \phi, \theta) = p(y | x, \phi) p(x | \theta),$$

where the functional forms of the two densities on the right-hand side are completely specified by the characterizing parameters $(\phi, \theta)$. The parameters $\phi$ and $\theta$ relate explicitly to the conditional for $y$ given $x$ and then the marginal for $x$, respectively. Though $\phi$ and $\theta$ are two distinct symbols in notation, they can be structurally dependent in various ways, as we will see later. From this joint density, we can also deduce the implied marginal density for $y$, $p(y|\phi, \theta)$, and the implied conditional density $p(x|y, \phi, \theta)$ via the complementary factorization

$$(2) \qquad p(y, x|\phi, \theta) = p(x|y, \phi, \theta)p(y|\phi, \theta),$$

where the full set of parameters $(\phi, \theta)$ may be involved, in complicated ways, in the "retrospective" conditional for $x$ given $y$, and the corresponding marginal for $y$.

The conditional density of $y$ given $x$ is essential for prediction, of course, and hence we center our development on the representation (1), in the knowledge that we can move interchangeably between factorizations (1) and (2) as desired.

## 2.2 Sampling Designs

The stochastic model of the data generation process, referred to as the sampling design, leads to likelihood functions as summaries of the data-based information on $(\phi, \theta)$. Typical sampling contexts fall into the following categories:

1. Data from the margins:

   - $Y^m = \{y_i^m, i = 1 : k_m\}$ where the $y_i^m \sim p(y|\phi, \theta)$ are independent, and/or
   - $X^m = \{x_i^m, i = (k_m + 1) : (k_m + n_m)\}$ where the $x_i^m \sim p(x|\theta)$ are independent,

   and with $Y^m$ and $X^m$ independent given $(\phi, \theta)$. Having the opportunity to observe data $Y^m$ provides information on aspects of the full set of parameters $(\phi, \theta)$, while $X^m$ informs on aspects of $\theta$ alone. $X^m$ is the traditional *unlabeled data*, though the same term could also be applied to $Y^m$.

2. Full prospective random sampling in which $(Y^p, X^p) = \{(y_i^p, x_i^p); i = 1 : n_p\}$ are drawn from the full joint distribution $p(y, x|\phi, \theta)$. Here data are paired and provide information on both $\theta$ and $\phi$. This is a common classification and/or regression design.

3. Data from a prospective design in which the $X^p = \{x_i^p, i = 1 : n_p\}$ values above are specified in advance by design. Then $X^p$ contains no information about the parameters and we learn about the parameter $\phi$ (only) through the likelihood based on $(Y^p, X^p)$ that is the product of components

$p(y_i^p|x_i^p, \phi)$—this is the venerable and perhaps the most common regression design in applied statistics. In machine learning the term transductive framework, outlined by Vapnik (1998), has been applied in this setting when the objective is to make predictions of $y$ on only some specific, prespecified values of $x$.

4. Data from a typical retrospective design—or case-control design—in which we observe the outcomes $X^r = \{x_i^r, i = 1 : n_r\}$ at a chosen set of $y$ values $Y^r = \{y_i^r, i = 1 : n_r\}$. Here too the data are paired, but $Y^r$ provides no information about $(\phi, \theta)$ since the $y$ values are chosen by design. The data in $X^r$ comprise a set of $n_r$ independent random draws from $p(x|y, \phi, \theta)$ and therefore provide information about $(\phi, \theta)$.

The difference between "prospective" and "retrospective" is whether the observed $y$ values are random or not. Since most examples we will discuss come from a prospective design, for notational simplicity we will drop the superscript and use $(Y, X)$ to denote $(Y^p, X^p)$. Other sampling schemes arise in statistical design (e.g., matched case-control designs, repeated measurement designs), but the above examples are key and central to much of predictive modeling and to our main goals of explicating the use of unlabeled data. Finally we note that the machine learning community has used the term "sampling" for a somewhat different use, applying it to different factorizations of the joint distribution assuming the data were generated by a full random sampling of $(y, x)$; in that usage, the form (1) is referred to as "diagnostic sampling" and (2) is referred to as "generative sampling" (Cozman and Cohen, 2002).

## 2.3 Prediction

We observe data $D$ generated via one or a combination of the sampling designs mentioned above. We aim to predict (estimate, classify) a new case $y_*$ at a value $x_*$. The prediction problem is solved from the Bayesian perspective by evaluating the posterior predictive distribution

$$(3) \qquad \begin{aligned} &p(y_*|x_*, D) \\ &= \iint p(y_*|x_*, \phi)p(\phi, \theta|x_*, D)\, d\phi\, d\theta \end{aligned}$$

at the value of the future $x_*$, where $p(\phi, \theta|x_*, D)$ is the posterior distribution of the parameters given the data and $x_*$. This posterior predictive distribution is the relevant quantity whether $x_*$ is a random draw from $p(x|\theta)$ or is specified directly. In the former case $x_*$ arises as a sample from $p(x|\theta)$ and so provides additional information about $\theta$; then $p(\phi, \theta|x_*, D)$ depends on $x_*$. In the latter case $x_*$ is chosen at a value of interest, often one of a range of values where we aim to explore potential future outcomes, and so provides no additional information; then

$$(4) \qquad p(\phi, \theta|x_*, D) = p(\phi, \theta|D).$$

In any example it is important to be aware of the distinction but, for our development, it is a side issue and we assume the latter case (4) as it simplifies the notation.

Our interest focuses on how $X^m$ enters in the evaluation of the predictive density in (3). All forms of information enter through $D$, so for $X^m$ (and any other information) to be relevant in prediction it is necessary that it play a role in defining the posterior $p(\phi, \theta|D)$. This is the key to understanding, if, and how, any information in $D$ impacts the prediction problem.

A relatively general framework has observations on each of $Y^m$, $X^m$, $(Y, X)$ and $(Y^r, X^r)$. Then Bayes' theorem under a specified prior $p(\phi, \theta)$ yields

$$p(\phi, \theta|D) \propto p(\phi, \theta)p(D|\phi, \theta)$$

with

$$p(D|\phi, \theta) = p(Y, X|\phi, \theta)p(X^m|\theta)$$
$$\cdot \, p(Y^m|\phi, \theta)p(X^r|Y^r, \phi, \theta).$$

This posterior will depend in complicated ways on all aspects of $D$, including aspects of the unlabeled data $X^m$. Investigating this dependence is the key to understanding the relevance and specific potential uses of unlabeled data.

## 2.4 Common Framework of Regression and Classification

For convenience and clarity, we start our discussion in the simple regression/classification context where data arise from a joint random sample $D = (X, Y)$. Then

$$p(\phi, \theta|D) \propto p(\phi, \theta)p(Y|X, \phi)p(X|\theta).$$

For example, we may have a linear or nonlinear regression model for $(y|x, \phi)$ in which $\phi$ represents the uncertain regression parameters or regression functions.

Now imagine that we have the opportunity to additionally observe or measure some unlabeled data $X^m$. The modified posterior with $D = \{Y, X, X^m\}$ is then

$$p(\phi, \theta|D) \propto p(\phi, \theta)p(Y|X, \phi)p(X|\theta)p(X^m|\theta).$$

If $\phi$ and $\theta$ are independent under the prior, $p(\phi, \theta) = p(\phi)p(\theta)$, then

$$p(\phi, \theta|D) = p(\phi|Y, X)p(\theta|X^m, X).$$

Thus, prior independence leads to posterior independence and the unlabeled data $X^m$ is irrelevant in learning about $\phi$, and hence irrelevant in predicting new $y_*$, if $\phi$ and $\theta$ are *a priori* independent. This follows from

$$p(y_*|x_*, D) = \iint p(y_*|x_*, \phi)p(\phi, \theta|D) \, d\phi \, d\theta$$
$$= \int p(y_*|x_*, \phi)p(\phi|Y, X) \, d\phi$$

by posterior independence.

In other cases, the posterior for $(\theta, \phi)$ may—and often will—involve dependencies. Therefore, additional information generated from marginal data will have an impact on the prediction problem via the integration over the posterior that defines $p(\phi, \theta|x_*, D)$. In the general framework, data from $Y^m$, $X^m$ and $(Y^r, X^r)$ will all have an impact on the prediction problem.

It is now evident that, beyond probabilistic/prior dependencies in the Bayesian formulation, any structural relationship between the "regression component" parameters $\phi$ and the "$x$-marginal" component parameters $\theta$ will also inevitably lead to dependence of predictions on the unlabeled data. How such dependencies arise and what forms they take depend on context.

A standard and conceptually simple example is that of mixture modeling, in which learning about the marginal distribution for $x$ informs on the relative probabilities of mixture components for the joint distribution for $(y, x)$, and hence influences predictions. This idea-fixing example is developed below as the first of a series of common modeling contexts that illuminate the issues and theoretical framework.

## 3. SOME CENTRAL MODELING CONTEXTS AND EXAMPLES

### 3.1 Nonlinear Regression Prediction Using Mixtures

A methodologically central example, and one in which the relevance of unlabeled data is transparent,

is that of Gaussian mixture modeling for regression. Consider the case of univariate $y$ and multivariate $x$, with joint sampling density

$$(5) \qquad p(y, x) = \sum_{i=1}^{m} \pi_i f_i(y, x),$$

where $0 \leq \pi_i \leq 1$, $\sum_i \pi_i = 1$ and the $f_i(y, x)$ are density functions of distinct multivariate normal distributions. Transforming the joint distribution (5) to the prospective parametrization in (1), we have

$$p(x | \theta) = \sum_{i=1}^{m} \pi_i f_i(x),$$

$$p(y | x, \phi) = \sum_{i=1}^{m} w_i(x) f_i(y | x),$$

where $f_i(x)$ and $f_i(y | x)$ are the corresponding marginal and conditional densities of the multivariate normal $f_i(y, x)$, and

$$w_i(x) = \frac{\pi_i f_i(x)}{\sum_j \pi_j f_j(x)}$$

is the conditional mixing probability evaluated at the conditioning $x$ value.

In terms of the general theory and notation of Section 2, the simplest parameter specification has $\phi = \{\pi, \alpha, \beta\}$ and $\theta = \{\pi, \alpha\}$, where $\pi = \{\pi_i : i = 1 : m\}$, $\beta$ is the full set of linear regression coefficients, intercepts and conditional variances in the set of $m$ normal linear models $f_i(y | x)$ and $\alpha$ is the full set of mean vectors and variance matrices of the normal distributions $f_i(x)$. This parametrization makes clear the direct structural dependence of $\phi$ and $\theta$, and hence the deductions from the general theory of Section 2 that unlabeled data will matter in future predictions. This conclusion is evident by inspection. Observing data on the margin $x$ provides direct information on the relative weights $\pi_i$ of the normal components, and hence provides information relevant to predicting future $y_*$ values. The unlabeled data also of course inform about the other parameters $\alpha$ of the margin for $x$, that is, the component mean and covariance parameters of the marginal normal mixture $p(x | \theta)$ as well as the component weights. These parameters are also involved in the calculations needed for prediction—through the conditional mixing probabilities $w_i(x_*)$—and so the unlabeled data play a more intricate role than just advising on the weights.

*An illustrative example.* A simple illustrative example fixes ideas and shows how unlabeled data may increase predictive accuracy in nonlinear regression via a mixture model. Consider two-dimensional data $(y, x)$ modeled using a three-component Gaussian mixture model in the above framework, setting $f_i(y, x)$ to be the bivariate normal $N(\mu_i, \Sigma_i)$, for each $i = 1 : 3$. From the following prior on model parameters, we simulated one set of parameters and, given those parameters, drew a sample of 175 observations from the resulting three-component Gaussian mixture. Figure 1(a1) is a scatter plot of the 175 observations.

The prior,

$$(\pi_1, \pi_2, \pi_3) \sim \mathrm{Dir}(1/3, 1/3, 1/3),$$

$$(\mu_i | \Sigma_i) \sim N(0, \tau \Sigma_i), \quad (i = 1, 2, 3) \text{ with } \tau = 0.2,$$

$$\Sigma_i \sim \mathrm{IW}(d, S_0),$$

$$(i = 1, 2, 3) \text{ with } d = 3 \text{ and } S_0 = (4/3) I_{2 \times 2},$$

was used for posterior and predictive analysis of subsets of this full data set. Here Dir denotes a Dirichlet distribution and IW an inverse Wishart, in standard notation.

The standard Gibbs sampler for mixture models (Lavine and West, 1992; West, 1992) delivers Monte Carlo approximations to posterior and predictive distributions. In particular, given posterior samples of all model parameters (including the latent mixture component indicators for each sample), the posterior mean of the regression curve can be approximated pointwise over a range of values $x_* = [a, b]$ to deliver the estimated regression function $\mathbb{E}(y_* | x_*, D)$ over this range. This is plotted in Figure 1(a1) for the case in which $D$ is the full set of 175 observations and $x_* = [-2, 2]$. The corresponding estimates of the predictive density functions $p(y_* | x_*, D)$ are plotted for three different values of $x_* = \{-1.5, 0, 2\}$ in Figure 1(a2). These regression and density curves can be viewed as the "gold standards" as they fully utilize all the available data.

Assume now that we can only measure $y$ values for $x$ in the range $[-1, 1]$. This leaves us with a smaller labeled data set and an unlabeled set $X^m$. We can fit the model to the labeled data only, or to the labeled and unlabeled data. The Gibbs sampler can be easily extended to treat the $y$ values for the unlabeled $X^m$ as missing data and draw the corresponding labels. Such analysis results in Monte Carlo estimates of regression curves and predictive densities that can be compared to those from the full data analysis: Figure 1(b) presents results using only the labeled data and Figure 1(c) presents results using both the labeled and the unlabeled data. Comparison of these graphs with Figure 1(a) strikingly illustrates the differences, and highlights the improvements in prediction that can be obtained by incorporating the unlabeled data.

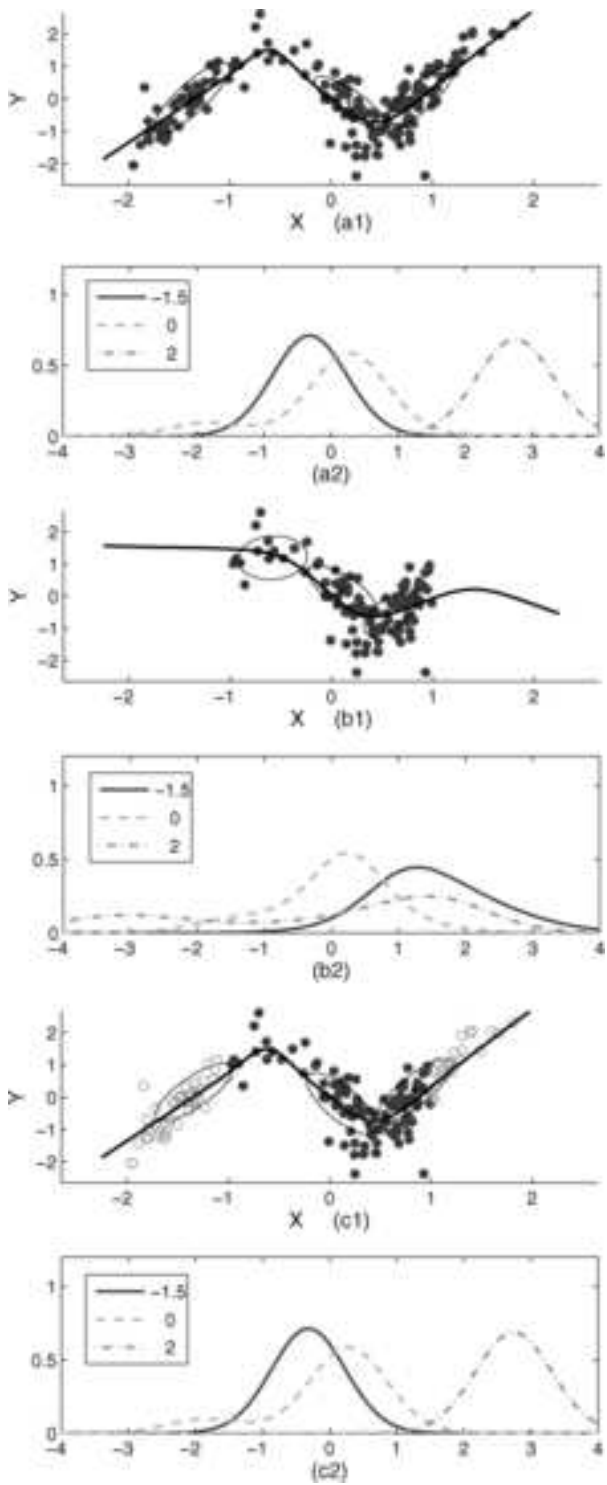

FIG. 1. *Mixture model regression and prediction example.* (a) *Analysis using the full data set:* (a1) *scatter plot of all 175 data points together with the posterior predictive regression curve of* $(y|x)$ *and three contours representing the posterior estimates of the three Gaussian components;* (a2) *the estimated predictive density functions* $p(y|x)$ *evaluated at three chosen values* $x = -1.5$, $0$ *and* $2$. (b) *Analysis using only the labeled data.* (c) *Analysis using the labeled data and also the unlabeled points (open circles).*

### 3.2 Classification and Discrimination with Mixtures

A related and also methodologically central mixture modeling context is that of classification and discrimination, in which $x$ arises from a mixture distribution and $y$ indicates the mixture component (Lavine and West, 1992). For example, in binary classification, for each of $y = 0, 1$ the model is specified with $\Pr(y = 1) = \pi$, $0 < \pi < 1$, and $(x|y) \sim f_y(x)$ for some parametrized densities $f_0$ and $f_1$. In the common Gaussian mixture model, $f_0$ and $f_1$ are multivariate normal densities parametrized by different means and variance matrices (Lavine and West, 1992), $f_y(x) = N(x|\mu_y, \Sigma_y)$ for each $y = 0, 1$. Define $\psi = \{\mu_0, \mu_1, \Sigma_0, \Sigma_1\}$. Here the scientifically natural specification of the joint distribution is via the "retrospective" construction of (2), parametrized as $p(x|y, \psi)$ and $p(y|\pi)$.

Relating to the factorization of (1), the implied marginal for $x$ and conditional for $y$ given $x$ are easily deduced; the former is the implied mixture of the two normal distributions weighted by probabilities $\pi$ and $1 - \pi$, and the latter is simply the revised outcome "classification" probability for $y = 1$ at a point $x$, computed by Bayes' theorem. It is transparent that unlabeled data matter in this setting. Observations from the marginal distribution for $x$ provide information about both the mixture weight $\pi$ and the parameters $\psi$ of the component normals. Prediction of a new $y_*$ at a point $x_*$ is performed by estimating (whether by formal Bayesian computations or otherwise) the classification probability $\Pr(y_* = 1|x_*)$, which is a complicated function of all parameters $\pi$ and $\psi$ and depends critically on various nonlinear functions of $\psi$ in particular. Hence information about $(\pi, \psi)$ from unlabeled data, as from any other source, feeds through to impact on predictions.

To connect with the general notation and theory of Section 2, we see that for the primary parameters $\pi, \psi$ there is no simple reduction or separation of the parameters into distinct parameters $\phi$ and $\theta$. Each of the distributions $p(y|x, \phi)$ and $p(x|\theta)$ depends in complicated ways on all the parameters $\pi$ and $\psi$, and consistency with the notation in Section 2 is achieved only by setting $\phi \equiv \theta = \{\pi, \psi\}$. Hence, in this key example, $\phi$ and $\theta$ are fundamentally highly structurally related, and the general theory of Section 2 implies that predictions will be impacted by the use of unlabeled data $X^m$ in concordance with our immediate, context-specific deductions above.

We note that various statistical and algorithmic approaches have been proposed to take advantage of the information in $X^m$ and the effectiveness of $X^m$ has been either implicitly or explicitly well studied methodologically (Ganesalingam and McLachlan, 1978; O'Neill, 1978; Ganesalingam and McLachlan, 1979; Müller, Erkanli and West, 1996; Nigam et al., 2000). An interesting theoretical connection in Castelli and Cover (1995) concerns the asymptotics of prediction errors for $y_*$ with respect to an increasingly large unlabeled sample, so that asymptotically all parameters are effectively "known"; the basic conclusion of this analysis was that labeled samples are exponentially more valuable than unlabeled samples in classification problems.

### 3.3 Normal Linear Regression Models

In the usual normal linear regression model, $\phi = (\beta, \tau)$ is the set of regression parameters from the model

$$y|x, \phi \sim N(\beta'x, \tau^2),$$

where $x$ and $\beta$ are $k$-dimensional vectors. One way such a model can arise is from an assumed joint multivariate normal distribution for $(y, x)$, namely, the $(k+1)$-dimensional (zero-mean) normal $N(0, \Sigma)$ where

$$\Sigma = \begin{pmatrix} \sigma_y^2 & \rho' \\ \rho & \Sigma_x \end{pmatrix},$$

for some scalar parameter $\sigma_y$, $k$-dimensional vector of covariance parameters $\rho$ and $k \times k$ variance matrix $\Sigma_x$. Under such a model we have $\beta = \Sigma_x^{-1}\rho$ and $\tau^2 = \sigma_y^2 - \beta'\rho$, and the marginal $p(x|\theta) = N(\mu_x, \Sigma_x)$ with the characterizing parameter $\theta = (\mu_x, \Sigma_x)$.

Some example contexts are as follows:

- A direct specification of the prior $p(\phi, \theta) = p(\phi) \cdot p(\theta)$ that assumes independence, and so implies that unlabeled $X^m$ data will be irrelevant to prediction of future $y_*$. This would be typical in many applied regression settings.

- An indirect specification in which the initial prior is defined for $(\mu, \Sigma)$, with the prior $p(\phi, \theta)$ being implied by transformation. A common approach is to use the conjugate normal-inverse Wishart prior distribution. Any prior in this class has the property that the implied prior on $(\phi, \theta)$ is in fact one in which $\phi$ and $\theta$ are independent (Geiger and Heckerman, 2002; Dobra et al., 2004).

- Other indirect specifications of the prior $p(\phi, \theta)$ by deduction from a prior on $\Sigma$ will induce dependence between $\phi$ and $\theta$ and hence lead to relevance of the unlabeled data since $X^m$ will then provide information about $\phi$ indirectly through its relevance for $\theta$.

The second example here illustrates a case in which modeling prior information on parameters of the joint distribution of $y$ and $x$ using a standard conjugate implies that the unlabeled $X^m$ data will be irrelevant for predicting $y_*$. This result arises more generally in exponential family models. Other priors may, and usually will, lead to prior and therefore posterior dependence so the unlabeled data will be relevant.

### 3.4 Binary Outcomes: Cancer Incidence and Prognosis

An illuminating example is the case of binary $y$ and binary $x$. For thematic context, suppose $x = 1/0$ represents the presence/absence of mutation in the BRCA1 breast cancer gene in a woman, and that $y = 1/0$ represents occurrence of breast cancer before age 70. The goal here is to predict the probability of $y = 1$ given the presence or absence of the mutation.

In this breast cancer example we define $\theta$ as the incidence rate of the BRCA1 mutation; $\phi_0$ is the base rate for breast cancer in the general (wild type) population of women, and $\phi_1$ the (higher) cancer rate among carriers of the mutation. The joint density using $p(y|x, \phi)$ and $p(x|\theta)$ is then parametrized by the three probabilities, $\phi = (\phi_0, \phi_1)$ and $\theta$ where

- $\phi_x = \Pr(y = 1|x, \phi)$ for $x = \{0, 1\}$ and
- $\theta = \Pr(x = 1|\theta)$.

For this model, the predictive distribution given the data is

$$p_* = \Pr(y_* = 1|x_*, D) = \iint \phi_{x_*} p(\phi, \theta|x_*, D) \, d\phi \, d\theta.$$

Given the above model, the following are two natural prior specifications:

- We can directly specify independent priors on $\theta$ and $\phi$. As a result $p_*$ will not depend on the unlabeled data.
- We have cell probabilities $p(x, y)$ on the joint space $x = 0, 1, y = 0, 1$ defined as

$$\pi = \{\pi_{00}, \pi_{01}, \pi_{10}, \pi_{11}\}.$$

Common approaches utilize Dirichlet priors on $\pi$. If we choose a Dirichlet prior $p(\pi)$ and find the implied prior $p(\phi, \theta)$ by transformation, the result is prior independence of $\phi$ and $\theta$, and again the unlabeled data are irrelevant to prediction.

A more interesting and perhaps natural modeling assumption on the joint space is that the breast cancer samples come from an inhomogeneous population having two genetically and environmentally different subpopulations in connection with inherited breast cancer-related characteristics and lifetime cancer risks. In this case a reasonable prior would be a mixture of two Dirichlets,

$$p(\pi) = ap_0(\pi) + (1 - a)p_1(\pi),$$

where $p_0$ and $p_1$ are two different Dirichlet priors for the two subpopulations, though the sampling design cannot distinguish between the subpopulations. It then follows by transformation that

$$p(\phi|\theta) = w(\theta)p_0(\phi) + (1 - w(\theta))p_1(\phi),$$

where $p_0$ and $p_1$ are the implied margins on $\phi$ from each of the two Dirichlets, and the mixing probability $w(\theta)$ is computed conditionally on any value of $\theta$ as

$$\frac{w(\theta)}{1 - w(\theta)} = \frac{a}{(1 - a)} \frac{p_0(\theta)}{(1 - p_1(\theta))}.$$

Thus under a mixture prior of this form, $\theta$ and $\phi$ are dependent and so the unlabeled data $X^m$ will provide information about $y_*$ indirectly via $\theta$ and $\phi$. The dependence between $\phi$ and $\theta$ is reflected in the variation of the weight $w(\theta)$ that provides the "link" for the unlabeled data information to flow through to impact on inferences about $\phi$, and hence to $y_*$. Only in the extreme case of no subpopulation structure, when $p_0(\cdot) = p_1(\cdot)$, will the unlabeled data on the mutational incidence rate play no role in predicting cancer events for future patients.

## 4. FACTOR MODELS AND FACTOR REGRESSION

### 4.1 Statistical Framework

The interest in factor regression has increased due to the prevalence of problems with high-dimensional predictors. One common example is principal component regression (PCR). In PCR, the singular value decomposition of the design matrix of original predictor variables generates principal components—or empirical factors—that become the predictors in a regression. The resulting orthogonal regression and potential data reduction are two key benefits of this modeling approach. However, a key question is raised in connection with prediction: since we aim to predict $y_*$ values at new, future $x_*$ values, should we not include the future design points in the initial analysis and principal component evaluation? This is evidently just a question of whether, and if so, how, to use unlabeled data in the model development and analysis of existing labeled data.

The question, and the general discussion of PCR and empirical factor regression, can be embedded in the broader theoretical context of (latent) factor regression models. West (2003) formalized the development of large-scale, latent factor models coupled with regression on latent factors, and delineated a comprehensive framework for predictive modeling that was particularly motivated by problems involving larger numbers of predictors—the "*large p, small n*" paradigm. This elucidated the theory underlying PCR and modeling using principal component projections of high-dimensional covariates/predictors as a limiting case of a broader class of regression models where the predictors are *latent* variables. This framework and theory also clarified and justified the use of so-called $g$-priors (Zellner, 1986) for Bayesian shrinkage regression, and defined novel classes of multiple shrinkage methods that are significantly beneficial in prediction problems through the ability to induce differential *shrinkage* in different factor–predictor dimensions. Importantly, the framework trivially clarifies the issue of use of unlabeled data, and how unlabeled samples enter into predictions based on analysis of labeled data, in general. The special limiting case of principal component regression is one important benefit.

The following normal linear model serves as a specific example to illustrate the more general principles of factor regression models. A univariate re-sponse $y$ is to be predicted based on a (high-dimensional) $p \times 1$ predictor variable $x$, and we have

$$y_i = \alpha' \lambda_i + \varepsilon_i \quad \text{and} \quad x_i = B\lambda_i + \nu_i,$$

where $\varepsilon_i \sim N(0, \sigma^2)$, $\lambda_i \sim N(0, I)$ is a $k \times 1$ multivariate normal latent factor for each $i$, $B$ is an uncertain $p \times k$ matrix of factor loadings of $x$ on $\lambda$, $\nu_i \sim N(0, \Psi)$ is a vector of idiosyncratic noise terms and $\Psi$ is an uncertain diagonal variance matrix. Also, the $\nu_i$ and $\varepsilon_i$ are conditionally (on all model parameters) mutually independent over $i$.

### 4.2 Unlabeled Data in Factor Regression Models

This framework is a key example of when unlabeled data matter. Fundamentally, the outcomes $y$ to be predicted are modeled as responses in regressions on *latent* variables $\lambda$, and the *observed* concomitant $x$ variables are related to $\lambda$, while $y$ and $x$ are conditionally independent *given* $\lambda$. Thus the predictive relevance of $x$ is indirect, through $\lambda$.

By marginalizing over $\lambda$ in the joint multivariate normal distribution of $y, x$ and $\lambda$ implied by the model specification, it becomes clear that we can identify $p(y|x, \phi)$ as a normal linear regression of $y$ on $x$ with regression parameter vector and residual variance making up the parameter $\phi = \phi(\alpha, \sigma, B, \Psi)$. Also, the implied marginal distribution for $x$ is normal with zero mean and variance matrix $\theta = BB' + \Psi$. Thus, if $\{B, \Psi\}$ are known, then $\theta$ is known and so the observed, unlabeled data $X^m$ has no influence whatsoever in the problem of predicting a future $y_*$ given $x$ data from either prospective or retrospective designs. However, typically $\{B, \Psi\}$ are uncertain and need to be estimated. In this setting:

- Unlabeled data $X^m$ provides information relevant to estimation of the latent factor model parameters $\{B, \Psi\}$, and hence of relevance to predicting future $y_*$ values via the transfer of information through inferences on the future $\lambda_*$ related to $x_*$.
- $\phi$ is dependent on aspects of $\theta$ indirectly through their functional associations with the factor model parameters, so that any relevant prior $p(B, \Psi, \alpha, \sigma)$ will induce dependencies between $\phi$ and $\theta$.

### 4.3 Digit Classification Example

The MNIST data set (Y. LeCun, http://yann.lecun.com/exdb/mnist/) is a standard data set used extensively in the machine learning community to benchmark binary regression models. The data set contains 60,000 images of handwritten digits $\{0, 1, 2, \ldots, 9\}$, where each image consists of $p = 28 \times 28 = 784$ gray-scale pixel intensities.

As an example, we consider what is generally regarded as one of the most difficult pairwise comparisons, that of discriminating a handwritten "6" from a "9." We frame this as a binary regression problem. The predictor space $x$ is transformed via singular value decomposition of the initial design matrix of 784 primary pixel values (after centering), and the first two factors are used for predictive discrimination of unlabeled samples.

The data set contains 5918 handwritten "6"s and 5949 handwritten "9"s. Following Belkin, Niyogi and Sindhwani (2004), we take the first 400 observations from each class as a training sample and use the remaining samples as test cases to be predicted. The standard MCMC analysis of the probit regression model produces approximate posterior predictive probabilities of "6" versus "9" for each of the several thousand test samples, and we record empirical prediction error rates based on whether or not the predictive probability of the true digit (true label) lies below or above 0.5. For the labeled/unlabeled evaluation, our analysis is most extreme: we randomly select just *two* "6"s and *two* "9"s to treat as labeled, the remaining 398 in each of the two classes being regarded as unlabeled. To give an initial indication of the relevance of unlabeled data, Figure 2 plots the projections of the full sets of training and test data onto the two factors (first two principal components) of the labeled and unlabeled data together; the separation of digits is quite strong and clear for both the training and test data. Repeating the factorization and projection of the training data, but now using only four labeled samples (randomly selected with two of each digit), produces the graph in Figure 3(a); frames (b), (c) and (d) of Figure 3 show similar plots for different random draws of the four samples treated as labeled. The relevance of unlabeled samples is quite evident from comparison of these plots with those based on the labeled and unlabeled data together.

In the probit factor regression models using the first two principal components as predictors, and with a simple, standard normal/inverse gamma prior on regression parameters (West, 2003), repeated analysis of labeled data alone in the above framework yields an average prediction error rate on the test samples of approximately 31.2%. Repeating this analysis but now including the unlabeled data in defining the empirical factors yields a semi-supervised average error rate of approximately 9.5%. This gives some indication of the potential improvements in raw predictive accuracy that may accrue from the appropriate use of unlabeled data.

# 5. KERNEL REGRESSION FOR PREDICTION AND CLASSIFICATION

## 5.1 Kernel Regression Models

An interesting class of examples, which is central to the methodological interfaces of statistics and machine learning, arises in models based on kernel regression. Kernel and related smoothing spline methods have a long history in applied statistics and have seen a tremendous amount of development at the interfaces of machine learning and statistics in the last several years (Poggio and Girosi, 1990; Wahba, 1990; Vapnik, 1998; Schölkopf and Smola, 2002; Shawe-Taylor and Cristianini, 2004; Liang et al., 2007).

The context is nonparametric, nonlinear regression with $y \in \mathbb{R}$, $x \in \mathbb{R}^k$, and a model of the form

$$(6) \qquad y = f(x) + \varepsilon,$$

where $\varepsilon$ is a zero-mean noise term and $f$ is an uncertain regression function. As an example, the class of Bayesian radial basis (RB) models (Liang et al., 2007) deals with questions of proper probability models—and the resulting proper inference and predictive results that then arise—for uncertain knots in a kernel model. This framework, and other approaches, begin with the interest in a representation of the form

$$(7) \qquad f(x) = \int w(u)K(x, u)\, dG(u)$$

for some weight function $w(u)$ over $k$-dimensional $u$, and some specified kernel function $K(\cdot, \cdot)$. The element $G(\cdot)$ is the unknown probability distribution function for $X$. The key to the model is to note that, if $G$ is discrete and puts masses $g_i$ at support points (or "knots") $u_i$, then the expression for $f(\cdot)$ is simply

$$f(x) = \sum_i g_i w(u_i)K(x, u_i),$$

that is, a radial basis function representation. The analysis of Liang et al. (2007) describes approximations to a model in which uncertainty about $G$ is expressed using a Dirichlet process prior (Ferguson, 1973; Escobar and West, 1995). One implication of such a model for $G$ is that, since Dirichlet processes are discrete with probability 1, the formal mathematical model for $f(x)$ is the sum above with a countably infinite number of knots $u_i$. From the methodological viewpoint, both labeled and unlabeled $x$ values provide information about $G$ directly. In fact, with a sample of $n$ labeled and/or unlabeled $x$ values $x_1, \ldots, x_n$ (whether from $X$, $X^m$ or some combination of the two), this Dirichlet process model implies that $f$ may be approximated by

$$(8) \qquad \hat{f}_n(x) = \sum_{i=1}^{n} w_{n,i}K(x, x_i),$$

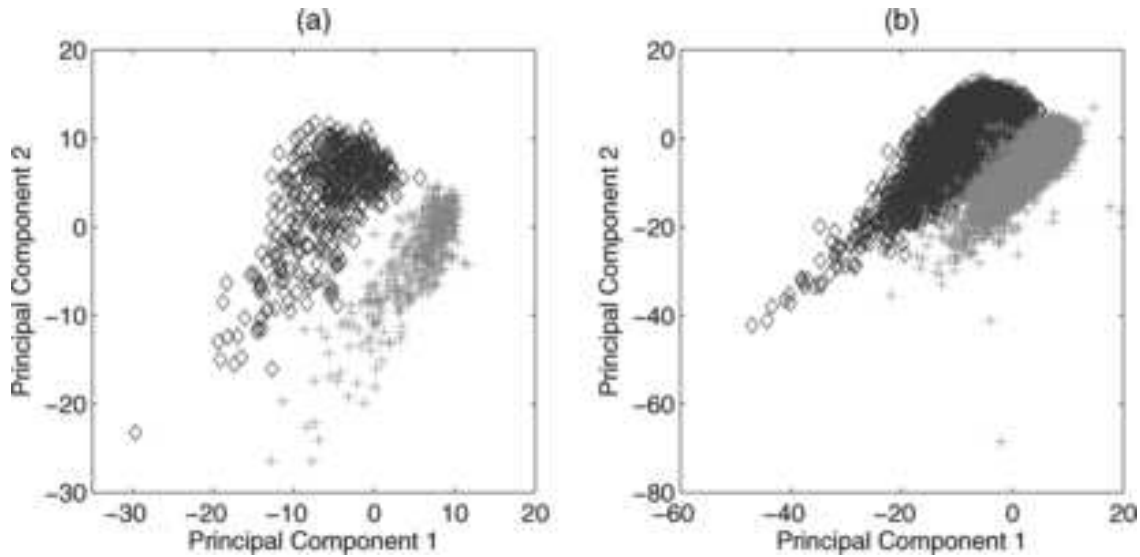

FIG. 2. *MNIST handwritten digit data example, with samples of handwritten "6" (+) and "9" (◇). The axes are the first two principal components computed from a singular value decomposition of the centered data, using the full data set of over 11,000 samples (5918 "6"s and 5949 "9"s). Scatter plotted on these empirical factors are* (a) *the training data of 800 samples (400 of each digit), and* (b) *the remaining test data. The two factors evidently carry strongly discriminating information.*

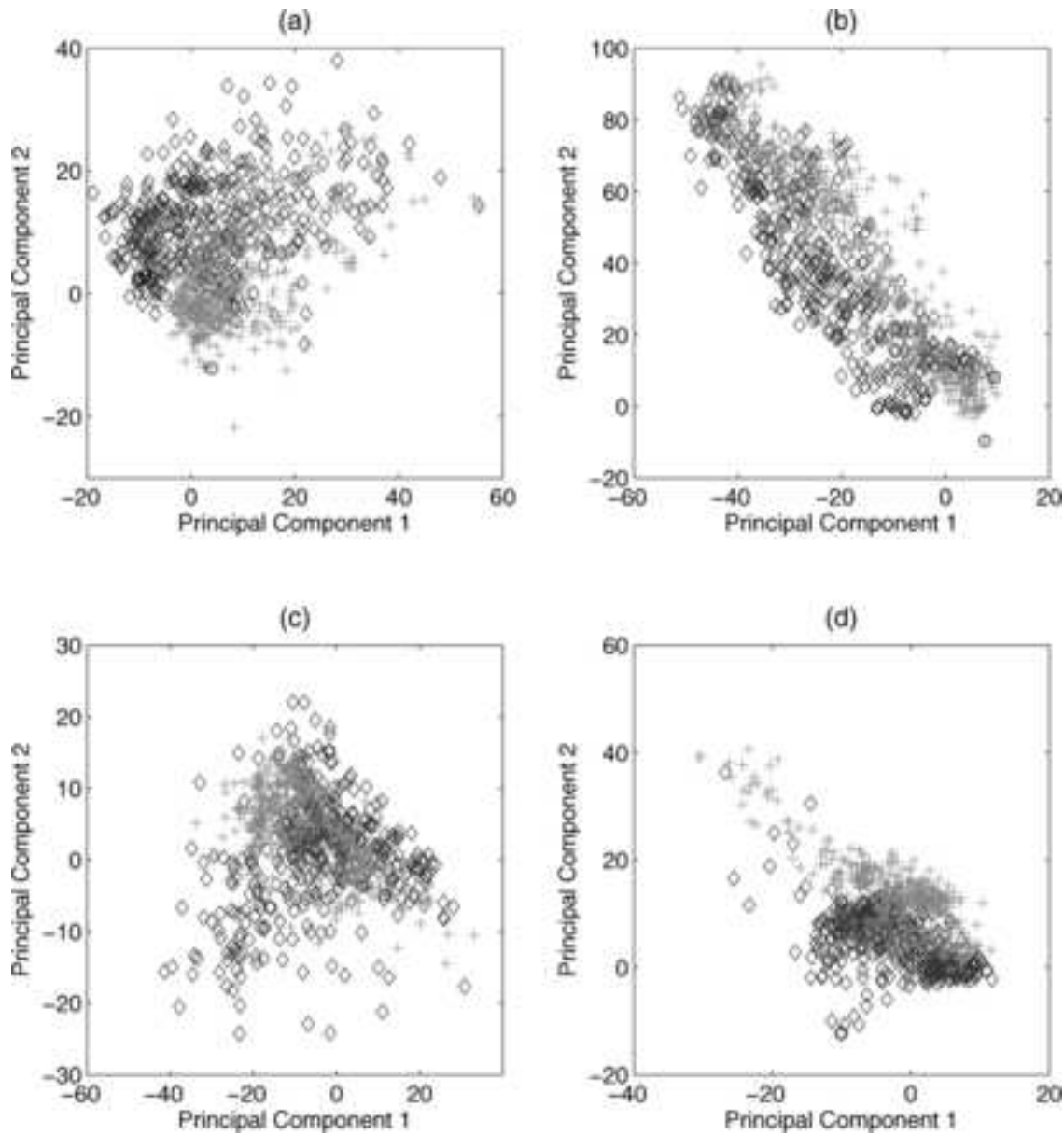

Fig. 3. *MNIST handwritten digit data example, in a format similar to that of Figure 2. In these frames the two principal components were evaluated on only four samples selected as labeled, two of each digit, and the scatter plots are of the training data of 800 samples. The four labeled samples were randomly drawn from the training data set, and the four frames here represent four different draws of the labeled samples. Though there is evidence of discriminatory information to distinguish the handwritten "6"s (+) from the "9"s (◇), it is quite clear that discriminatory power will be very limited.*

where $w_{n,i} \propto w(x_i)$. The key methodological relevance of this approach is that this is true for all $n$, providing consistency as sample size increases and additional design points are observed. This leads to the practical model in which each $y_*$ is linearly regressed on the set of kernel predictors $\{K(x_*, x_i)\}_{i=1}^n$ based on whatever set of design points is observed. A complete model now involves a prior distribution over the induced regression coefficients $w_{n,i}$ and we note that this explicitly depends on $n$ and the realized $x_i$. Hence, in both the structure of the regression model and in the requirements for a prior over coefficients, we see the dependence on all values observed in the $x$ space; this is therefore a perfect example of when, why and how unlabeled $X^m$ data matters. In particular, we note that:

• $\theta = G(\cdot)$ so that $p(x|\theta) \, dx = dG(x)$—the parameter is the full distribution function itself.

- Equations (6) and (7) show explicitly how $p(y|x, \phi)$ depends intimately on $\theta = G$ as defining the nonlinear kernel regression; in fact, $\theta \subseteq \phi$ in this case. Thus prior and posterior dependence of $\theta$ and $\phi$ is central to the model.
- As a result, unlabeled $X^m$ provides direct, immediate and critically relevant information in predicting $y_*$.

This specific example of a kernel regression model derived within a coherent probabilistic framework, taken from Liang et al. (2007), is presented for its simplicity and also because it represents a fully specified probabilistic model in which the kernel weights $w_{n,i}$ are related coherently as sample sizes change. Some additional connections and related kernel regression formulations are now mentioned.

## 5.2 Relation to Machine Learning Kernel Regression Algorithms

Other constructions of kernel regression models, including those utilizing Gaussian processes and spline smoothing, non-Bayesian uses of radial basis functions and others (Poggio and Girosi, 1990; Wahba, 1990; Schölkopf and Smola, 2002; Vapnik, 1998; Shawe-Taylor and Cristianini, 2004), exhibit the same structure and consequent dependence on unlabeled data. One interesting connection with recent theoretical developments in machine learning approaches arises by noting that the central model of (7) also corresponds to the solution of the nonlinear manifold regularization formulation of Belkin, Niyogi and Sindhwani (2004). This approach, motivated by geometric arguments, is an optimization algorithm that minimizes

$$f_* = \underset{f \in \mathcal{H}_K}{\arg\min} \left[ \frac{1}{n} \sum_{i=1}^n V(f(x_i), y_i) + \gamma_A \|f\|_K^2 + \gamma_I \|f\|_I^2 \right],$$

where $\{(y_i, x_i)\}_{i=1}^n$ are the labeled data, $\mathcal{H}_K$ is a reproducing kernel Hilbert space (RKHS), $V(f(x), y)$ is a loss function, $\|f\|_K^2$ is the RKHS norm, $\gamma_A, \gamma_I$ are regularization parameters and $\|f\|_I^2$ is a norm that reflects the smoothness of the function on the marginal $p(x)$. If the marginal is concentrated on a manifold, $x \subset \mathcal{M} \in \mathbb{R}^k$, then a natural choice for $\|f\|_I^2$ is the Laplacian on the manifold. The marginal $p(x)$ is generally unknown; with unlabeled data $X^m$ from the marginal, the Laplacian on the manifold

may be approximated by a Laplacian on the graph defined by the observed data (labeled and unlabeled)

$$\hat{f}_n(x) = \underset{f \in \mathcal{H}_K}{\arg\min} \left[ \frac{1}{n} \sum_{i=1}^n V(f(x_i), y_i) + \gamma_A \|f\|_K^2 + \frac{\gamma_I}{(n+n_m)^2} \mathbf{f}^T \mathbf{L} \mathbf{f} \right],$$

where $\mathbf{L}$ is the graph Laplacian on all the data (given a weight matrix on the graph) and $\mathbf{f} = \{f(x_1), \ldots, f(x_n), f(x_1^m), \ldots, f(x_{n_m}^m)\}$. The above optimization is achieved by

$$\hat{f}(x) = \sum_{i=1}^n w_{n,i} K(x, x_i) + \sum_{i=1}^{n_m} w_{n+n_m, n+i} K(x, x_i^m),$$

which takes the same form as (8). This formulation as an optimization problem from a statistical machine learning viewpoint generates precisely the same functional form of the model as that derived from a nonparametric regression in the Bayesian framework above, and the consequences for the use of unlabeled data in model formulation are the same.

## 5.3 Illuminating the Potential Impact of Unlabeled Data

A simple but illuminating synthetic example provides an initial illustration. A data set of 50 points $\{(x_i, y_i), i = 1:50\}$ is plotted in Figure 4(b); here $x_i \in \mathbb{R}^2$ and $y_i = 0/1$. This data set can be easily classified according to $y = 0$ versus $y = 1$ by a Gaussian kernel model, and we fit such a model using the Bayesian model completion—in terms of prior specification for the kernel weights and observational variance parameters—and the resulting MCMC method for model fitting as described in Liang et al. (2007). Though the details of the prior specification and computation are not central here, we note that the model involves use of a generalized shrinkage prior, termed *generalized g-prior* by West (2003), on the kernel regression coefficients. This is a method of importance when dealing with large numbers of regression parameters, and its use in these kernel models where the number of regression parameters exceeds the number of labeled observations is particularly apt.

The analysis leads to the computation of posterior predictive probabilities for $y = 1$ versus $y = 0$ at any chosen new $x$ value, that is, the class predictions based on any data set. Using the fully labeled 50 data points for such an analysis yields results displayed and described in Figure 4(a).

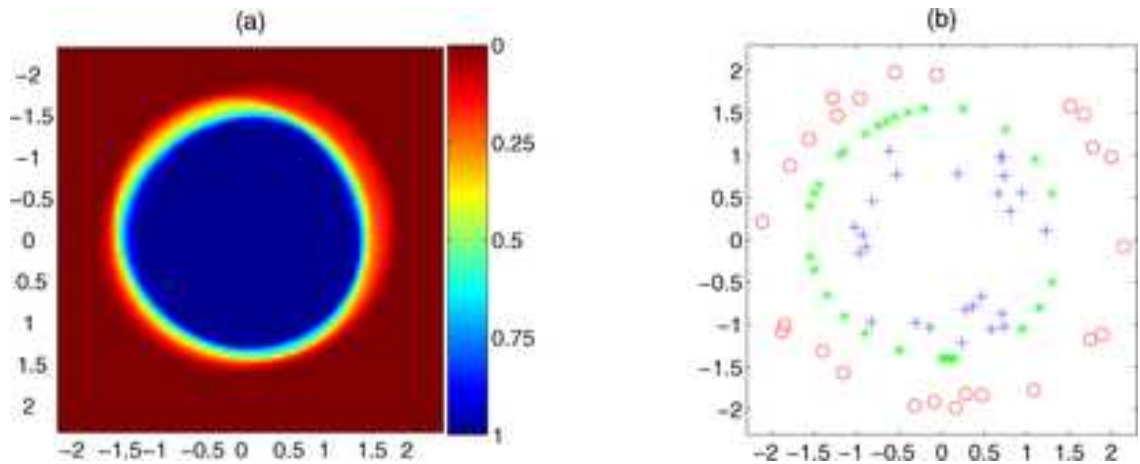

FIG. 4.   *Kernel regression example using synthetic data. Frame* (b) *displays a scatter plot of the 50 observations on the $x_2$ versus $x_1$ axes, with cases $y_i = 1$ as "+" (blue) and $y_i = 0$ as "o" (red). The binary kernel regression model analysis of the full set of 50 labeled observations produces approximate posterior predictive probabilities $\Pr(y = 1 | x, D)$ at any point $x$ in the plane; the green "*" points in frame* (b) *are points at which $\Pr(y = 1 | x, D) = 0.5$, that is, represent points on the separating contour. Frame* (a) *displays a color image of the contours of $\Pr(y = 1 | x, D)$ as $x$ varies; red corresponds to the conditional probability being near 0 and blue near 1.*

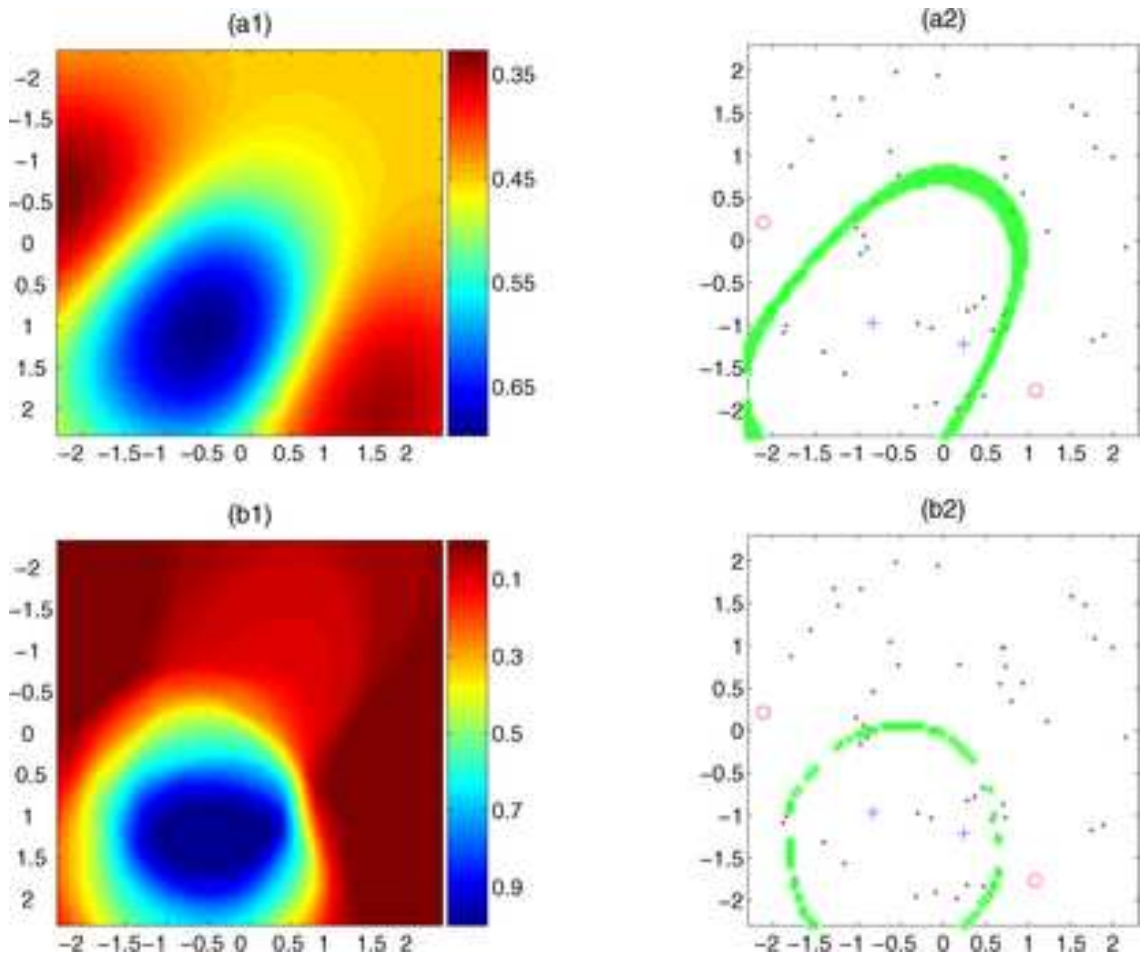

FIG. 5.   *Kernel regression example with displays as in Figure 4:* (a) *using only four selected data points;* (b) *using the same four selected data points but also including the unlabeled $X^m$ data.*

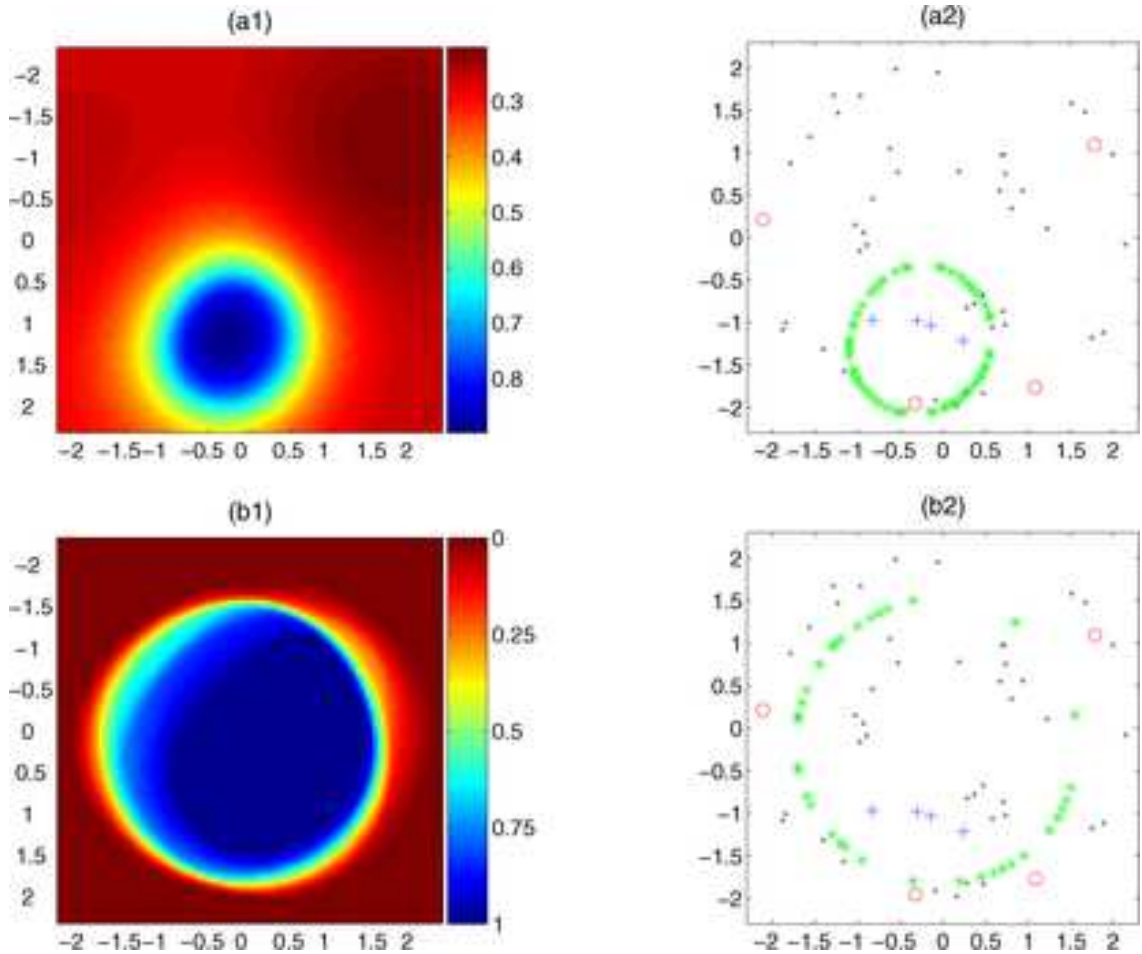

Fig. 6. *Kernel regression example: (a) using eight labeled data points; (b) using the eight labeled data points together with the remaining unlabeled samples. Notice that the use of the 42 unlabeled samples is sufficient to produce predictive probability contours that are very similar to those using the full set of labeled observations, as in Figure 4, though with evident and justifiably greater uncertainty, even though the y values are labeled on only eight data points.*

The analysis is repeated using only four labeled points, two each with $y = 0$ and $y = 1$; the four randomly selected points are marked in Figure 5. The resulting class prediction contours and summaries of predictions for the 46 unlabeled points are then computed in two separate analyses: (i) using only the labeled data—just the four points; and (ii) using the labeled and unlabeled data. Figure 5 presents the results of these two analyses. This exercise was repeated using a total of eight labeled points, resulting in the displays in Figure 6. From the figures the major impact of unlabeled data is clearly apparent, and its relevance with very small numbers of labeled samples highlighted. We also see that the semisupervised analysis using only eight labeled samples results in predictions that are very similar to those obtained if all 50 samples were labeled. Unlabeled data can dramatically impact upon and improve prediction accuracy.

## 5.4 Kernel Regression for Cancer Classification Using Genomic Data

A substantive example involves analysis of a gene expression data set consisting of DNA microarray expression profiles from 190 tissue samples representing a variety of different primary tumors (breast, prostate, lung, lymphoma, etc.) and 90 noncancerous, "normal" samples from the corresponding tissue of origin (Ramaswamy et al., 2001; Mukherjee et al., 2003). Following standard processes of data normalization and screening for genes showing nontrivial variation, the data analyzed consists of $p = 2800$ gene expression variables, or "genes," on the set of 280 samples. The analysis setup aims to use the gene expression data as predictors $(x)$ in a binary kernel regression model with outcome $y = 1$ representing "cancer," that is, any of the cancer types, and $y = 0$ representing normals. As in the synthetic example above, the analysis uses a Gaussian kernel model fitted using Bayesian shrinkage priors, as in Liang et al. (2007). Our interest is to compare predictions of cancer versus normal under this given model and prior specification applied to differing selections of data—selections that allow us to examine the impact of samples of data.

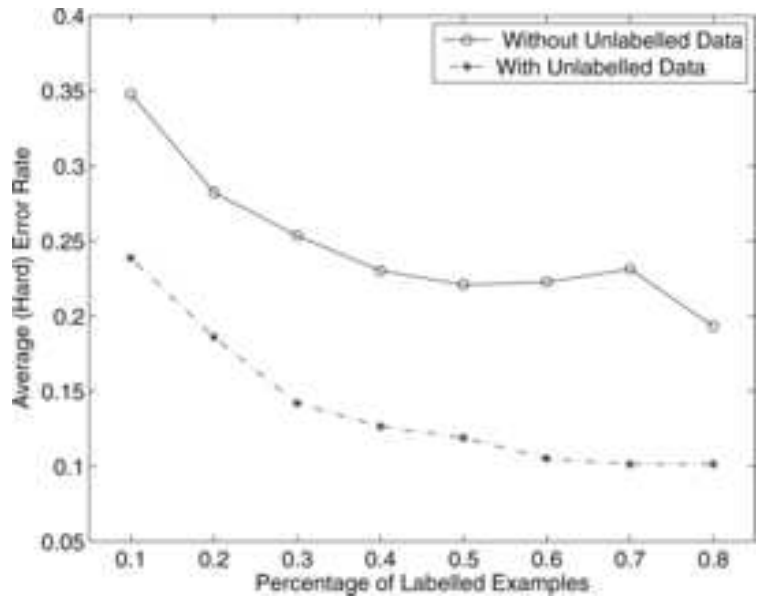

FIG. 7. *Results from predictions of cancer versus normal tissue ($y = 0, 1$) based on gene expression data ($x$), showing empirical prediction error rates from the analysis ignoring the unlabeled data (solid line) compared to those from the analysis including the unlabeled data (dashed line). These results were developed by repeating the analyses with varying percentages of the data (horizontal axis) randomly designated as unlabeled. The selection of unlabeled cases and model analysis was rerun 50 times for each chosen unlabeled percentage. The graph represents the average prediction error in predicting the true status ($y = 1$ versus $y = 0$ with predictive probability thresholded at 0.5). The uniform improvement in empirical predictive accuracy when including the unlabeled data is clear.*

We do this as in the synthetic example above—randomly selecting a fraction of the data to be regarded as unlabeled, fitting the model and then predicting the status (cancer versus normal) of the selected unlabeled cases in terms of posterior predictive probabilities. We repeat this analysis twice—first, using only the labeled data; second, using both the labeled and unlabeled data—and are then able to compare predictions between the two analyses to assess the changes due to use of the unlabeled data. For a given fraction of unlabeled data, we repeated this 50 times, each time randomly selecting the cases to be labeled/unlabeled, and computing the average (across the 50 repeats) empirical prediction error rate in classifying the unlabeled cases. The prediction of an unlabeled sample is regarded as "correct" if the predictive probability of the true state (cancer or normal) exceeds 0.5. Figure 7 summarizes the resulting empirical error rates for a series of such analyses in which we progressively increased the percentage of labeled data from 10% to 90%. The figure clearly shows the differences between analysis using only labeled data and that using labeled and unlabeled data.

Additional insight into the impact of including unlabeled samples is given in Figure 8. From one analysis with 80% of the data unlabeled, we select 10 each of the cancer and normal samples that were unlabeled in the analysis, and graph the estimated predictive probabilities of cancer versus normal with approximate 95% credible interval. This shows the impact of the unlabeled $x$ data on the predictions, in terms of the impact on estimates of prediction uncertainty as well as empirical accuracy.

## 6. SUMMARY COMMENTS

Beginning with an articulation of the basic sampling and design specifications underlying statistical formulations of prediction problems, we have delineated the conceptual and theoretical issues underlying the use and relevance, or irrelevance, of unla-beled data in classification and prediction problems. This, coupled with a series of examples in central statistical modeling contexts, and empirical illustrations and evaluations in two substantive data analyses, provides an overview and synthesis of the ideas underlying the emerging methodology of semisupervised learning in the machine learning and statistics literatures.

Graphical model representations of the joint sampling model context aid in this interpretation. The relevance, or otherwise, of the unlabeled $X^m$ data can be deduced essentially by inspection of the implied (undirected) graphical representation of any full model structure. For example, the full distribution assuming joint sampling, and in cases for which $p(\phi, \theta) = p(\phi)p(\theta)$, is illustrated in graphical terms in Figure 9. The joint density exhibited here is

$$p(y_*, x_*, Y, X, X^m, \phi, \theta)$$
$$= p(y_* | x_*, \phi)p(Y | X, \phi)p(X | \theta)p(x_* | \theta)$$

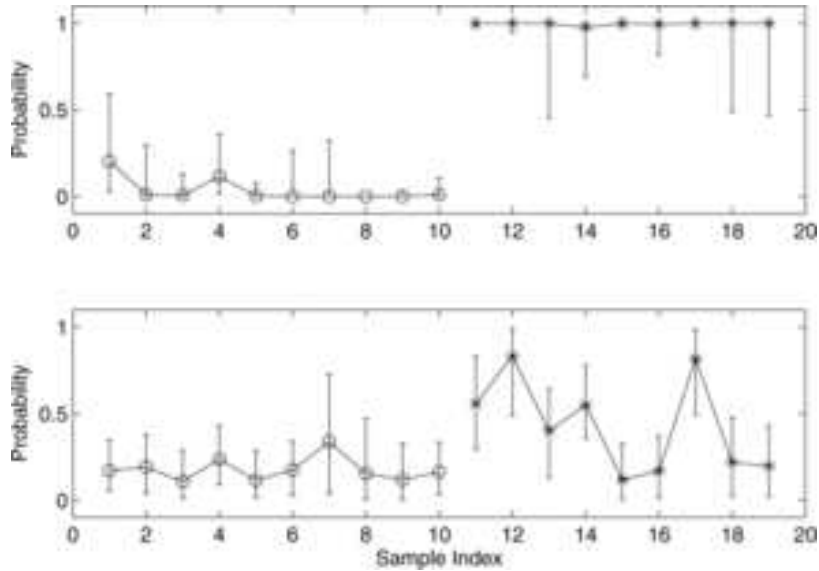

FIG. 8. *Cancer versus normal predictions using kernel model. The figure displays estimated predictive probabilities of cancer versus normal for 10 cancers (∗) and 10 normal tissues (○) that were unlabeled in the data analysis. This analysis involved only 20% of the data being labeled. The frames also provide estimated 95% credible intervals associated with each of the predictions. This shows the impact of the unlabeled x data on the predictions, in terms of the impact on estimates of prediction uncertainty as well as empirical accuracy.*

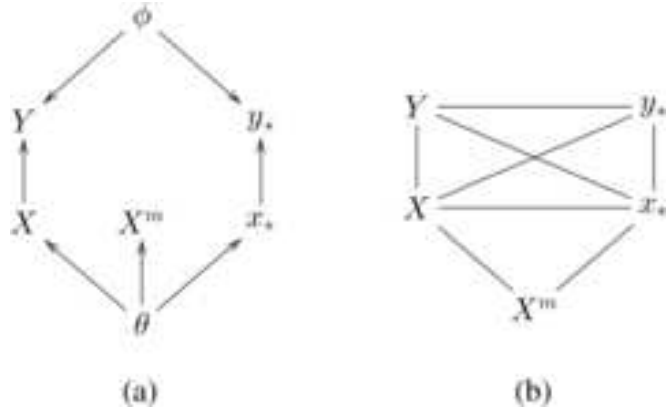

FIG. 9. *Graphical models of the basic structure relevant to understanding the role of unlabeled data in predictive modeling. The figure shows the directed (acyclic) graph (a) and undirected graph (b) of the joint distribution of data and parameters in cases of independence of $\phi, \theta$. In contrast, if $\phi, \theta$ are dependent, then (a) would have an edge between $\phi$ and $\theta$, and (b) would be a fully connected graph.*

$$\cdot \, p(X^m|\theta)p(\phi)p(\theta).$$

Figure 9(a) is a directed acyclic graph of the joint distribution structured in terms of composition of sampling distributions. Figure 9(b) displays the corresponding undirected graph in which the lack of an edge between $X^m$ and $y_*$ indicates conditional independence given all other quantities, hence the irrelevance to prediction of the unlabeled data in this case. In contrast, were $\phi, \theta$ to be *a priori* dependent, then the five nodes of the undirected graph would be fully connected, exhibiting the relevance of the unlabeled data to prediction of $y_*$.

In addition to clarifying and exemplifying the structure of models and the prediction problem with unlabeled data, one aim of this work has been to review the area to provide a link across the mainstream statistical and machine learning communities. We hope that this will entice more statistical researchers into a very active, productive and exciting research milieu, while also founding the discussion in venerable, simple and unambiguous terms arising from the direct and classical probabilistic formulation. This view directly, we believe, addresses and answers the questions of "when, why and how" unlabeled data help in predictive modeling.

## ACKNOWLEDGMENTS

The authors acknowledge the most constructive suggestions of the Executive Editor and an anonymous Editor on the original version of this paper. The research reported here was partially supported by National Science Foundation grants DMS-03-42172 and DMS-04-06115. Any opinions, findings and conclusions or recommendations expressed in this work are those of the authors and do not necessarily reflect the views of the NSF.